\documentclass[conference]{IEEEtran}
\usepackage{cite}
\usepackage{url}
\usepackage[dvipsnames,svgnames]{xcolor}
\usepackage{graphicx}
\usepackage[tight,footnotesize]{subfigure}
\usepackage{algorithmic}
\usepackage{overpic}
\usepackage{amssymb}
\usepackage{psfrag}
\usepackage{hyperref}
\DeclareGraphicsExtensions{.xps}
\usepackage{epstopdf} 
\usepackage[font=small,labelfont=bf]{caption}
\usepackage[cmex10]{amsmath}
\usepackage{amssymb}
\usepackage{amsthm}

\newtheorem{theorem}{Theorem}

\def\CN{\mathcal{N}_{\mathbb{C}}} 
\newcommand{\vect}[1]{\mathbf{#1}}
\def\tr{\mathrm{tr}}

\def\tr{\mathrm{tr}}

\def\Htran{\mbox{\tiny $\mathrm{H}$}}

\def\CN{\mathcal{N}_{\mathbb{C}}} 
\def\taupu{\tau_{p}} 

\begin{document}
\IEEEoverridecommandlockouts
\title{Deep Learning Power Allocation in Massive MIMO}
\author{
\IEEEauthorblockN{Luca Sanguinetti\IEEEauthorrefmark{2}\IEEEauthorrefmark{3}, Alessio Zappone\IEEEauthorrefmark{3}, Merouane Debbah\IEEEauthorrefmark{3}\IEEEauthorrefmark{4}\bigskip
\thanks{\hrulefill \newline The work of L. Sanguinetti work was supported by the University of Pisa under the PRA 2018-2019 Research Project CONCEPT and by the H2020-ERC PoC-CacheMire project (grant 727682). The research of A. Zappone was supported by the H2020 MSCA IF BESMART, grant 749336. The work of M. Debbah was partly supported by the H2020 MSCA IF BESMART, grant 749336, and by the H2020-ERC PoC-CacheMire project, grant 727682.
}}
\IEEEauthorblockA{\IEEEauthorrefmark{2}\small{Dipartimento di Ingegneria dell'Informazione, University of Pisa, Pisa, Italy}}
\IEEEauthorblockA{\IEEEauthorrefmark{3}\small{Large Networks and System Group (LANEAS), CentraleSup\'elec, Universit\'e Paris-Saclay, Gif-sur-Yvette, France}}
\IEEEauthorblockA{\IEEEauthorrefmark{4}\small{Mathematical and Algorithmic Sciences Lab, Huawei Technologies, France.}}

\vspace{-.8cm}}
\maketitle

\begin{abstract}
This work advocates the use of deep learning to perform max-min and max-prod power allocation in the downlink of Massive MIMO networks. More precisely, a deep neural network is trained to learn the map between the positions of user equipments (UEs) and the optimal power allocation policies, and then used to predict the power allocation profiles for a new set of UEs' positions. The use of deep learning significantly improves the complexity-performance trade-off of power allocation, compared to traditional optimization-oriented methods. Particularly, the proposed approach does not require the computation of any statistical average, which would be instead necessary by using standard methods, and is able to guarantee near-optimal performance. 
\end{abstract}



\section{Introduction}\label{Sec:Intro}
Massive MIMO refers to a wireless network technology where the base stations (BSs) are equipped with a very large number $M$ of antennas to serve a multitude of user equipments (UEs) by spatial multiplexing \cite{Marzetta2010a,Larsson2014}. Exciting developments have occurred in the recent year. In industry, the technology has been integrated into the 5G New Radio standard \cite{Parkvall2017a}. In academia, the long-standing pilot contamination issue, which was believed to impose fundamental limitations \cite{Marzetta2010a}, has finally been resolved \cite{BHS18A}. More precisely,  \cite{BHS18A} showed that with optimal minimum mean squared error (MMSE) combining/precoding and a tiny amount of spatial channel correlation, the capacity increases without bound in uplink (UL) and downlink (DL) as the number of antennas increases. 

In this work, we propose to use deep learning for solving the max-min and max-prod power allocation problems in the DL of Massive MIMO networks. We are inspired by the recent explosion of successful applications of machine learning techniques \cite{Bengio2016} that demonstrate the ability of deep neural networks to learn rich patterns and to approximate arbitrary function mappings \cite{Bengio2016,Hornik89}. Particularly, we aim to demonstrate that the positions of the UEs (which can be easily obtained via global positioning system) can be effectively used by a neural network to obtain near-optimum performance. This allows to reduce substantially the complexity of power allocation (since simple matrix-vector operations are required) and thus makes it possible to perform power allocation in real-time, i.e. following the variations of UEs' positions. In addition to this, training such a neural network is fairly convenient since training samples are easily obtainable by running off-the-shelf optimization algorithms. 

Deep learning for radio resource allocation in wireless networks has been also considered in \cite{SunDL}, where the WMMSE algorithm for sum-rate maximization has been emulated by a fully-connected feedforward neural network, and in \cite{YuDL2018}, where a convolutional neural network is used for user-cell association.


\section{Massive MIMO network}\label{Sec:SystemModel}
We consider the DL of a Massive MIMO network with $L$ cells, each comprising a BS with $M$ antennas and $K$ UEs \cite{massivemimobook}.
We denote by $\vect{h}_{li}^{j} \in \mathbb{C}^{M}$ the channel between UE~$i$ in cell~$l$ and BS~$j$ and assume that
\begin{equation} \label{eq:correlated-Rayleigh-model}
\vect{h}_{li}^{j} \sim \CN \left( \vect{0}_{M}, \vect{R}_{li}^{j}  \right)
\end{equation}
where $\vect{R}_{li}^{j} \in \mathbb{C}^{M \times M}$ is the spatial correlation matrix, known at the BS. The normalized trace $\beta_{li}^{j} = {1}/{M} \tr ( \vect{R}_{li}^{j})$
accounts for the average channel gain from an antenna at BS~$j$ to UE~$i$ in cell~$l$ and is modelled as (in dB)
\begin{align}\label{pathloss}
\beta_{li}^j = \Upsilon - 10 \alpha \, \log_{10} \left( \frac{d_{li}^{j}}{1\,\textrm{km}} \right)\quad \text{dB}
\end{align}
where $\Upsilon = -148$ dB determines the median channel gain at a reference distance of 1\,km, and $\alpha=3.76$ is the pathloss coefficient. Also, $d_{li}^{j}$ is the distance of UE $i$ in cell $l$ from BS $j$, given by $d_{li}^{j} = \|{\bf{x}}_{li}^{j}\|$
with ${\bf{x}}_{li}^{j}\in \mathbb{R}^2$ being the UE location in the Euclidean space. Note that shadowing should also be considered in \eqref{pathloss}. However, this is usually modeled by a log-normal distribution, resulting into a channel model that is not spatially consistent. In other words, two UEs at almost the same location would not experience the same channel. To overcome this issue, one should resort to channel models based on ray tracing or recorded measurements.

\subsection{Channel Estimation}
Pilot-based channel training is utilized to estimate the channel vectors at BS $j$. We assume that the BS and UEs are perfectly synchronized and operate according to a time-division duplex (TDD) protocol wherein the DL data transmission phase is preceded in the UL by a training phase for channel estimation. There are $\taupu=K$ pilots (i.e., pilot reuse factor of $1$) and UE $i$ in each cell uses the same pilot. Using a total UL pilot power of $\rho^{\rm{tr}}$ per UE and standard MMSE estimation techniques \cite{massivemimobook}, BS $j$ obtains the estimate of $\vect{h}_{li}^j$ as
\begin{align}
\hat{\vect{h}}_{li}^j = \vect{R}_{li}^j \vect{Q}_{li}^{-1} \bigg( \sum_{l'=1}^{L} \vect{h}_{l'i}^j +\frac{1}{\tau_p} \frac{\sigma^2}{\rho} \vect{n}_{li}   \bigg) \!\sim \!\CN \left( \vect{0},  \vect{\Phi}_{li}^j \right)
\end{align}
where $\vect{n}_{li} \sim \CN (\vect{0}, \vect{I}_{M})$ is noise, $\vect{Q}_{li} = \sum_{l'=1}^{L} \vect{R}_{l'i}^j + \frac{1}{\rho^{\rm{tr}}} \vect{I}_{M}$, and $
\vect{\Phi}_{jli}  = \vect{R}_{li}^j \vect{Q}_{li}^{-1} \vect{R}_{li}^j$.
The estimation error $\tilde{\vect{h}}_{li}^j = \vect{h}_{li}^j - \hat{\vect{h}}_{li}^j  \sim \CN ( \vect{0}, \vect{R}_{li}^j- \vect{\Phi}_{li}^j)$ is independent of $\hat{\vect{h}}_{li}^j$.

\subsection{Downlink Spectral Efficiency}
The BS in cell~$l$ transmits the DL signal $
\vect{x}_l = \sum_{i=1}^{K} \vect{w}_{li} \varsigma_{li}$
where  $\varsigma_{li} \sim \CN(0,\rho_{li})$ is the DL data signal intended for UE~$i$ in cell $l$, assigned to a precoding vector $ \vect{w}_{li} \in \mathbb{C}^{M}$ that determines the spatial directivity of the transmission and satisfies $ \|  \vect{w}_{li} \|^2  =1$ so that $\rho_{li}$ represents the transmit power. 

An achievable DL SE can be computed in Massive MIMO by using the following \emph{hardening bound} \cite{Marzetta2016a}. 
\begin{theorem} \label{theorem:downlink-capacity-forgetbound} 
The DL ergodic channel capacity of UE~$k$ in cell~$j$ is lower bounded by 
\begin{equation} \label{eq:downlink-SE-expression-forgetbound}
{\mathsf{SE}}^{\mathrm{dl}}_{jk} = \frac{\tau_d}{\tau_c} \log_2   (  1 +
\gamma^{\mathrm{dl}}_{jk} ) \quad \textnormal{[bit/s/Hz]} 
\end{equation}
with 
\begin{align}\label{eq:downlink-SINR-expression-forgetbound}
\!\!\!\gamma^{\mathrm{dl}}_{jk} =  \frac{ \rho_{jk}| \mathbb{E}\{ \vect{w}_{jk}^{\Htran} \vect{h}_{jk}^{j} \} |^2  }{ 
\sum\limits_{l=1}^{L} \sum\limits_{i=1}^{K}  \rho_{li}\mathbb{E} \{  | \vect{w}_{li}^{\Htran} \vect{h}_{jk}^{l}  |^2 \}
- \rho_{jk}| \mathbb{E}\{ \vect{w}_{jk}^{\Htran} \vect{h}_{jk}^{j} \} |^2 + {\sigma^2}  } 
\end{align}
where the expectations are computed with respect to the channel realizations. The pre-log factor $\frac{\tau_d}{\tau_c}$ accounts for the fraction of samples per
coherence block used for DL data.
\end{theorem}

Notice that the above lower bound is achieved when the UE treats the mean of its precoded channel as the true one.
This is a reasonable assumption for channels that exhibit \emph{channel hardening} \cite[Sec. 2.5]{massivemimobook}, but a certain loss occurs for channels
with little or no hardening. An alternative approach (not considered in this work) consists in estimating the precoded
channels either explicitly as in \cite{Ngo2017a} or implicitly as in \cite{Caire17}.

\subsection{Precoder Design}
Unlike in the UL \cite[Sec. 4.1]{massivemimobook}, finding the optimal precoders is a challenging task since the DL SE in \eqref{eq:downlink-SINR-expression-forgetbound} depends on the precoding vectors $\{\vect{w}_{li}\}$ of all UEs in the entire network. Motivated by the UL-DL duality \cite[Sec. 4.3]{massivemimobook}, a common heuristic approach is to select $\vect{w}_{jk}$ as
\begin{equation} \label{eq:precoding-based-on-combining}
\vect{w}_{jk} = \frac{ \vect{v}_{jk} }{ \|  \vect{v}_{jk} \| } 
\end{equation}
where $\vect{v}_{jk}$ denotes the combining vector used to detect the UL signal transmitted by UE $k$ in cell $j$. In this work, we assume that $\vect{v}_{jk}$ is designed according to MR combining \cite{Marzetta2016a}
\begin{equation} \label{eq:precoding-schemes}
\vect{v}_{j k}^{\rm MR}=\hat{\vect{h}}_{j k}^j
\end{equation}
and M-MMSE combining \cite{BHS18A,Li2016a}
\begin{equation} \label{eq:precoding-schemes}
\vect{v}_{j k}^{\rm M-MMSE} = \Bigg(  \sum\limits_{l=1}^L \sum\limits_{i=1}^K \hat{\vect{h}}_{li}^j {(\hat{\vect{h}}_{li}^j)}^{\Htran} + \vect{Z}_j  \Bigg)^{\!-1}  \!\!\!  \hat{\vect{h}}_{jk}^j
\end{equation}
where $\vect{Z}_j = \sum\nolimits_{l=1}^{L} \sum\nolimits_{i=1}^{K} (\vect{R}_{li}^j - \vect{\Phi}_{li}^j) +  \frac{\sigma_{\rm {ul}}^2}{\rho_{\rm {ul}}}  \vect{I}_{M}$. This choice is motivated by the fact that M-MMSE is optimal but has high computational complexity. On the other hand, MR is suboptimal (not only for finite values of $M$ but also as $M\to \infty$ \cite{BHS18A}) but has the lowest complexity among the receive combining schemes.

\section{Power Allocation}

The DL SE of UE~$k$ in cell~$j$ can be rewritten as
\begin{equation} \label{eq:downlink-SE-expression-forgetbound}
{\mathsf{SE}}^{\mathrm{dl}}_{jk} = \frac{\tau_d}{\tau_c} \log_2   \Bigg(  1 +  \frac{ \rho_{jk} a_{jk}  }{ 
\sum\limits_{l=1}^{L} \sum\limits_{i=1}^{K}  \rho_{li} b_{lijk} + \sigma^2  }  \Bigg) \quad \forall j,k
\end{equation}
where \vspace{-0.2cm}
\begin{align}\label{ajk}
a_{jk} = | \mathbb{E}\{ \vect{w}_{jk}^{\Htran} \vect{h}_{jk}^{j} \} |^2 
\end{align} 
and
\begin{align} \label{b_lijk}
\!\!b_{lijk} = \begin{cases} \mathbb{E} \{  | \vect{w}_{jk}^{\Htran} \vect{h}_{jk}^{l}  |^2 \} & \!\!\!(l,i) \neq (j,k) \\
\mathbb{E} \{  | \vect{w}_{li}^{\Htran} \vect{h}_{jk}^{l}  |^2 \} - | \mathbb{E}\{ \vect{w}_{li}^{\Htran} \vect{h}_{jk}^{j} \} |^2 & \!\!\!(l,i) = (j,k)
\end{cases} 
\end{align} 
are the average channel gains and average interference gains, respectively. The average is computed with respect to the small-scale fading realizations so that the DL SE is only a function of the large scale fading statistics and the choice of precoding. This is a unique feature of Massive MIMO that largely simplifies the power allocation problem compared to single-antenna systems \cite[Sec. 7.1]{massivemimobook}. 

Among the different power allocation policies, two prominent examples are the max-min fairness and max product SINR strategies, which can be mathematically formalized as follows: 
\begin{align} \label{eq:max-min}
\max_{\{\rho_{jk}:\forall j,k\}} & \quad \, \min_{j,k} \mathsf{SE}_{jk}^{\rm dl} \\
\textrm{subject to}\,\,\,\,\, & \quad  \sum_{k=1}^{K} \rho_{jk} \leq P_{\max}^{\mathrm{dl}} , \quad j=1,\ldots,L \notag
\end{align}
and the max product SINR, given by
\begin{align} \label{eq:max-prod}
\max_{\{\rho_{jk}:\forall j,k\}} & \quad \, \prod_{j=1}^{L} \prod_{k=1}^{K} \gamma_{jk} ^{\rm dl} \\
\textrm{subject to}\,\,\,\,\, & \quad  \sum_{k=1}^{K} \rho_{jk} \leq P_{\max}^{\mathrm{dl}} , \quad j=1,\ldots,L \notag
\end{align}
where $P_{\max}^{\mathrm{dl}}$ denotes the maximum DL transmit power. Irrespective of the strategy, the following Monte Carlo methodology is needed to compute the optimal powers \cite{massivemimobook}.

\begin{enumerate}
\item Macroscopic propagation effects
\begin{enumerate}
\item Randomly drop UEs in positions ${\bf{x}}_{li}^{j}$
\item Compute large-scale fading coefficients $\beta_{lk}^j$
\item Compute channel correlation matrices $\vect{R}_{lk}^j$
\end{enumerate} 
\item Microscopic propagation effects
\begin{enumerate}
\item Generate random estimated channel vectors $\hat{\vect{h}}_{lk}^j$ by using MMSE estimator
\end{enumerate} 
\item SE computation \begin{enumerate}
\item Compute precoding vectors $\vect{w}_{jk}$ with MR or M-MMSE precoding
\item Average over estimated channels to obtain $\{a_{jk} \}$ and $\{b_{lijk}\}$.
\end{enumerate}
\item Allocate the power by solving \eqref{eq:max-min} or \eqref{eq:max-prod}.
\end{enumerate}
The solution to \eqref{eq:max-min} can be obtained through a bisection approach in which a sequence of convex problems is solved, while  \eqref{eq:max-prod} can be solved by geometric programming. Thus, both \eqref{eq:max-min} and \eqref{eq:max-prod} require a polynomial or quasi-polynomial complexity to be solved. However, even a polynomial complexity can be too much when the solution must be obtained in real-time; that is, fast enough to be deployed in the system before the UEs' positions change and the power allocation problem needs to be solved again. 

\section{Deep Learning based Power Allocation}
A central goal of this work is to demonstrate that geographical location information of UEs is already sufficient as a proxy for computing the optimal powers at any given cell. This is in contrast to the traditional optimization approaches for solving \eqref{eq:max-min} and \eqref{eq:max-prod} that require knowledge of $\{a_{jk} \}$ and $\{b_{lijk}\}$ in \eqref{ajk} and \eqref{b_lijk}. We advocate using UEs' positions because they already capture the main feature of propagation channels and interference in the network. Therefore, for any given cell $j$ the problem is to learn the \emph{unknown} map between the solution ${{\boldsymbol{\rho}}}_{j}^{\star}=[\rho_{j1}^{\star},\ldots,\rho_{jK}^{\star}]\in\mathbb{R}^{K}$ to \eqref{eq:max-min} or \eqref{eq:max-prod} and the $2KL$ geographical UE positions ${\bf x} = \{{\bf{x}}_{li}^{j}; \forall j,l,i\}\in\mathbb{R}^{2KL}$. This is achieved by leveraging the known property of NNs that are universal function approximators \cite{Hornik89,Bengio2016}.
Particularly, we employ a feedforward neural network with fully-connected layers, 
and consisting of a $2KL$-dimensional input layer, $N$ hidden layers, and a $K+1$-dimensional output layer yielding an estimate ${\hat {\boldsymbol{\rho}}}_{j}=[\hat {\rho}_{j1},\ldots,\hat {\rho}_{jK}]$ of the optimal power allocation vector ${{\boldsymbol{\rho}}}_{j}^{\star}$. Observe that the output layer has size $K+1$ instead of $K$, since we also make the NN learn $\sum_{k=1}^{K} \rho_{jk}^{\star}$ so as to satisfy the power constraint and increase the estimation accuracy.

The problem reduces to train the weights $\bf W$ and bias terms $\bf b$ of the NN so that the input-output map of the NN emulates the map of traditional approaches. This requires a training set containing $N_T$ multiple samples $\{{{\boldsymbol{\rho}}}_{j}^{\star}(n),{\bf x}(n);n=1,\ldots,N_T\}$, where ${{\boldsymbol{\rho}}}_{j}^{\star}(n)$ corresponds to the optimal power allocation  for the training input ${\bf x}(n)$. 
Denoting by ${\hat {\boldsymbol{\rho}}}_{j}(n)$ the corresponding output of the NN, the learning process consists of minimizing the following loss:
\begin{equation}
\displaystyle \min_{{\bf W},{\bf b}}\;\frac{1}{N_T}\sum_{n=1}^{N_T}\ell({\hat {\boldsymbol{\rho}}}_{j}(n),{{\boldsymbol{\rho}}}_{j}^{\star}(n))
\end{equation}
with $\ell(\cdot,\cdot)$ any suitable distance measure. Once the parameters ${\bf W}$ and ${\bf b}$ are configured, the NN can estimate the optimal power allocation policy also for input vectors that are not part of the training set. Therefore, every time the UEs' change their positions in the network, the power allocation can be updated by simply feeding the new positions to the NN, without having to actually solve \eqref{eq:max-min} or \eqref{eq:max-prod}. The complexity reduction granted by this approach is analyzed in more detail in the next section.

\subsection{Online implementation and complexity}
The complexity of the proposed approach mainly lies in the generation of the training set. Assume that each layer is composed of $N_{i}$ neurons. Computing the output of the NN requires only $\sum_{i=1}^{N+1}N_{i-1}N_{i}$ real multiplications\footnote{The complexity related to additions is neglected as it is much smaller than that required for multiplications.} and the evaluation of $\sum_{i=1}^{N+1}N_{i}$ activation functions. Also, the training algorithm is conveniently performed by standard (stochastic) gradient descent algorithms coupled with the \emph{back-propagation} algorithm \cite[Ch. 6.5]{Bengio2016}. Instead, generating the training set requires to actually solve  \eqref{eq:max-min} or \eqref{eq:max-prod}
for many different realizations of ${\bf x}$, by means of traditional optimization theory methods. However, this is not an issue for at least two reasons:
\begin{itemize}
\item The training set can be generated \emph{off-line}. Thus, a much higher complexity can be afforded and real-time constraints do not apply.
\item The training set can be updated at a much longer time-scale than the rate at which the UEs' positions in the network vary. Thus, the training set can be updated at a much longer time-scale than that at which the power control problem should be solved if traditional resource allocation approaches were used.
\end{itemize}
From the above considerations, it follows that the proposed approach grants a huge complexity reduction, which allows one to update the power allocation based on the UEs' positions in real time.

\begin{table}[t]\label{network_parameters}
\centering
\begin{tabular}{|c|c|}
\hline
      Cell area (with wrap-around) &  $1$\,km $ \times\,  1$\,km \\
Bandwidth & $20$\,MHz  \\
Number of cells &  $L = 4$ \\
Number of UEs per cell&  $K=5$ \\
UL noise power & $ -94$\,dBm \\
UL transmit power & $20$\,dBm\\
Samples per coherence block & $\tau_c = 200$ \\
Pilot reuse factor & $1$\\
\hline
\end{tabular}
\caption{Massive MIMO network.}\label{table1}
\end{table}

\section{Performance evaluation}
We consider the Massive MIMO network reported in Table 1 with $L=4$ cells, with each cell covering a square area of $250\times 250$ m. A wrap around topology is used. We assume that $K=5$ UEs are randomly and
uniformly distributed in each cell, at distances larger than 35\,m from the BS. Results are averaged over 100 UE distributions. We consider communication over a 20\,MHz bandwidth with a total receiver noise power $\sigma^2$ of $-94$\,dBm. We assume that $\tau_p =K$ (i.e., pilot reuse factor of $1$) and that the UL transmit power $\rho$ per UE is $20$ dBm.  

The NNs were trained based on a dataset of $N_T =340000$ samples of independent realizations of the UEs' positions $\{{\bf x}(n);n=1,\ldots,N_T\}$, and optimal power allocations $\{{{\boldsymbol{\rho}}}_{j}^{\star}(n);n=1,\ldots,N_T\}$ for $j=1\ldots,L$, obtained by solving \eqref{eq:max-min} and \eqref{eq:max-prod} with traditional optimization approaches. Particularly, 90\% percent of the samples was used for training and 10\% for validation. Other $10000$ samples formed the test dataset, which is independent from the training dataset. The training set is available online at \href{https://data.ieeemlc.org}{https://data.ieeemlc.org} while the Matlab code available at \href{https://github.com/lucasanguinetti/}{https://github.com/lucasanguinetti/}
allows to generate further samples. We used the Adam optimizer \cite{KingmaB14}, and chose the relative MSE as loss function $\ell(\cdot,\cdot)$ since numerical results showed that it performed better than the MSE for the problem at hand. The learning rate, batch size, and epochs were adjusted with a trial and error approach. We used the open source python library Keras. The code is available online at \href{https://github.com/lucasanguinetti/}{https://github.com/lucasanguinetti/}.

{{\begin{table}[t]
\centering
\begin{tabular}{c|c|c|c}
   & Size& Parameters& Activation function\\
\hline

Input & 40 &--& --\\
Layer 1 (Dense) & 64 &2624 & elu\\
Layer 2 (Dense) & 32 &2080 &elu\\
Layer 3 (Dense)& 32 &1056&elu\\
Layer 4 (Dense)& 32 &528&elu\\
Layer 5 (Dense)& 5 &85&elu\\
Layer 6 (Dense)& 6 &36&linear
\end{tabular}\caption{Layout of the neural network. The trainable parameters are $6,373$.}\label{table1}
\end{table}}}

\subsection{Max-prod}
To evaluate the performance of the NN-based power allocation, we illustrate the cumulative distribution function (CDF) of the DL SE per UE, where
the randomness is due to the UE locations and shadow fading realizations. We consider MR, and M-MMSE. The NN used with both precoding schemes is reported in Table \ref{table1}, whose trainable parameters
are $6,373$. The results of Fig. \ref{fig1a} show that the NN matches very well the optimal solution with M-MMSE. The average MSE is $0.007$. With MR precoding, a small mismatch between the two curves is observed. Indeed, the average MSE increases to $0.051$. Fig. \ref{fig1b} illustrates the CDF of the  MSE of the SEs. As expected, the CDF curve with M-MMSE is to the left of the MR curve. This basically means that the NN achieves, statistically speaking, better performance with M-MMSE than with MR. This result might seem counterintuitive, since the M-MMSE is algorithmically and computationally more complex than MR and thus its optimal power allocation should in principle be more difficult to learn. A possible explanation for this is that with MR precoding the power is allocated only on the basis of the desired signal gain. On the other hand, with M-MMSE this is accomplished by also taking into account the power of interfering signals. Since the NN receives as input the positions of all UEs in the network, it is able to make the most of this information only when M-MMSE is employed. 

To improve the learning capabilities with MR, we also considered the more complex NN reported in Table \ref{table2}. Numerical results show that the average MSE of SEs reduces to $0.003$ and $0.015$ with M-MMSE and MR precoding, respectively. This is achieved at the price of an computational complexity and training time since the number of trainable parameters
is $202,373$, instead of $6,373$. 

To conclude, with the max-prod strategy the proposed deep learning based power allocation has significant computational complexity advantage compared to traditional approaches, while maintaining near-optimal performance with both MR and M-MMSE precoding.

\begin{figure}[t!]
\begin{center}
\subfigure[CDF of the DL SE per UE]{\label{fig1a}
\begin{overpic}[unit=1mm,width=1\columnwidth]{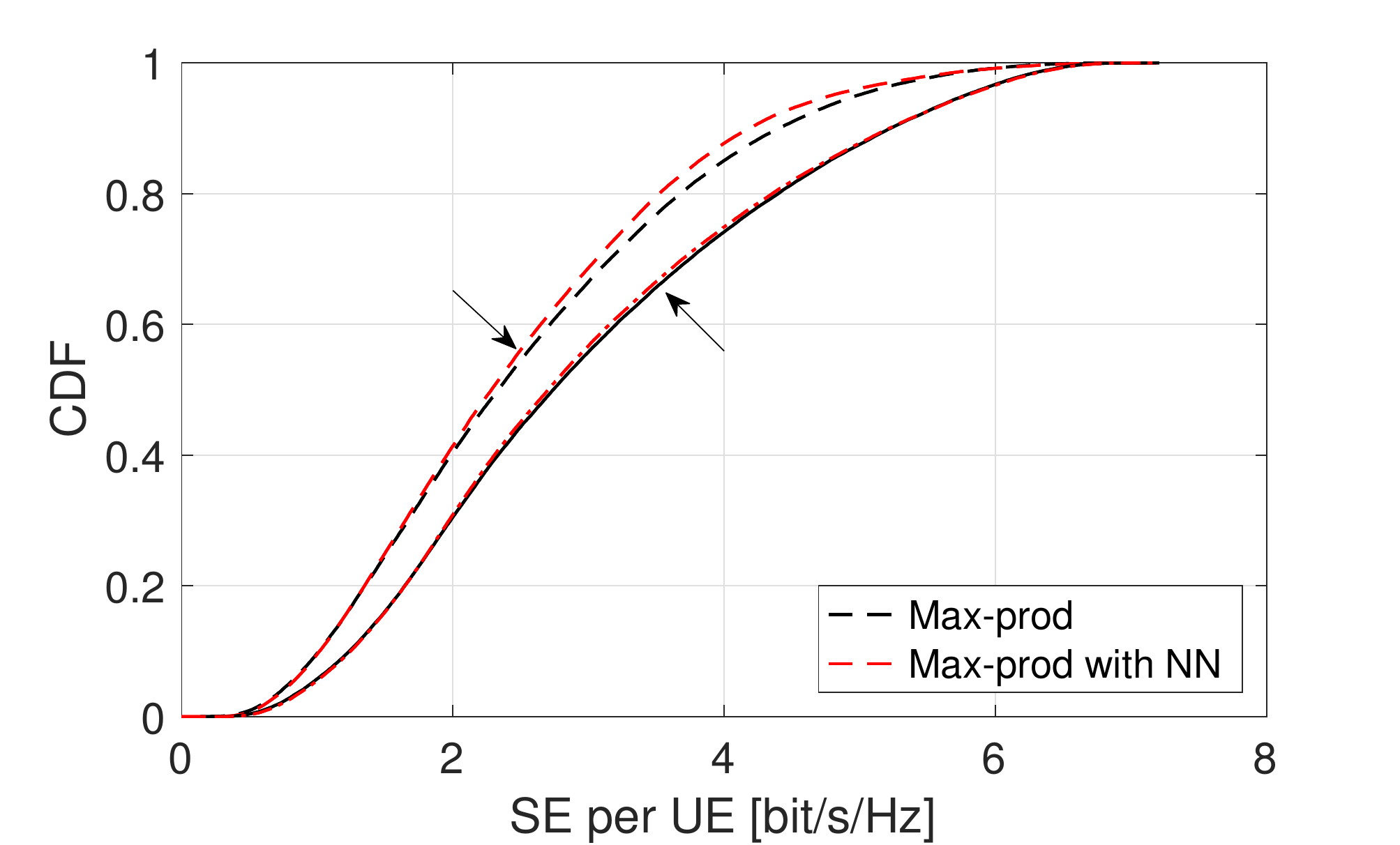}
\put(27,41){\small{MR}}
\put(48,32){\small{M-MMSE}}
\end{overpic}} 
\subfigure[CDF of the MSE of SEs]{\label{fig1b}
\begin{overpic}[unit=1mm,width=1\columnwidth]{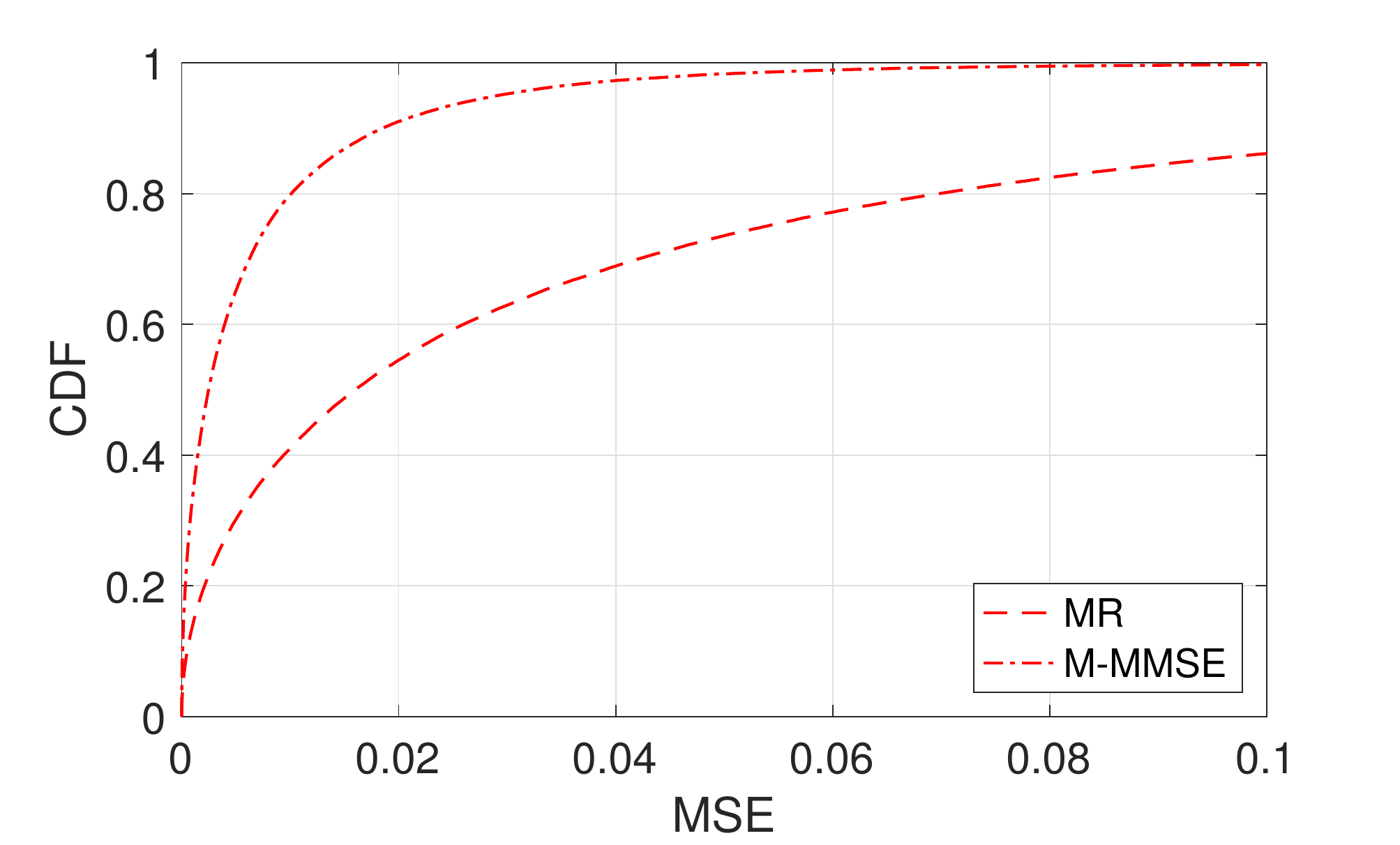}
\end{overpic}} 
\end{center} 
\caption{CDF of the DL SE and of the MSE of SEs with MR and M-MMSE precoding by using the NN of Table \ref{table1}.}\label{SE_with_imperfect_statistics}
\end{figure}

{\begin{table}[t]
\renewcommand{\arraystretch}{1.}
\centering
\begin{tabular}{c|c|c|c}
   & Size& Parameters& Activation function\\
\hline

Input & 40 &--& --\\
Layer 1 (Dense) & 512 &20992 & elu\\
Layer 2 (Dense) & 256 &131328 &elu\\
Layer 3 (Dense)& 128 &32896&elu\\
Layer 4 (Dense)& 128 &16512&elu\\
Layer 5 (Dense)& 5 &645&elu\\
Layer 6 (Dense)& 6 &36&linear
\end{tabular}
\caption{Layout for a given cell with $L=4$ and $K=5$. Trainable params: 202,373}\label{table2}
\end{table}}

{\begin{table}[t]
\renewcommand{\arraystretch}{1.}
\centering
\begin{tabular}{c|c|c|c}
   & Size& Parameters& Activation function\\
\hline

Input & 40 &--& --\\
Layer 1 (LSTM) & 256 &204128 & tanh\\
Layer 2 (LSTM) & 128 &197120 &tanh\\
Layer 3 (Dense)& 64 &8256&relu\\
Layer 4 (Dense)& 5 &325&relu
\end{tabular}
\caption{Layout for a given cell with $L=4$ and $K=5$. Trainable params: 509,829}\label{table4}
\end{table}}

\subsection{Max-min}
The NNs used for the max-prod strategy, revealed to be inadequate with the max-min approach. This is probably due to the fact that the power distribution changes considerably between the two strategies. To overcome this issue, we used a different NN, which consists of two recurrent Long Short-Term Memory (LSTM)1 layers and two dense layers. The NN parameters together with the activation functions are summarized in Table \ref{table4}. The results of Fig. \ref{SE_with_imperfect_statistics} show that the NN matches almost exactly the theoretical curves with both MR and M-MMSE. Despite providing satisfactory results in terms of accuracy, the NN in Table \ref{table4} counts a total number of $509,829$ trainable parameters. This is a relatively high number for a Massive MIMO network with $L=4$ and $K=5$. It lacks scalability when the network size increases.

\begin{figure}[t!]
\begin{center}
\subfigure[CDF of the DL SE per UE]{\label{fig1a}
\begin{overpic}[unit=1mm,width=1\columnwidth]{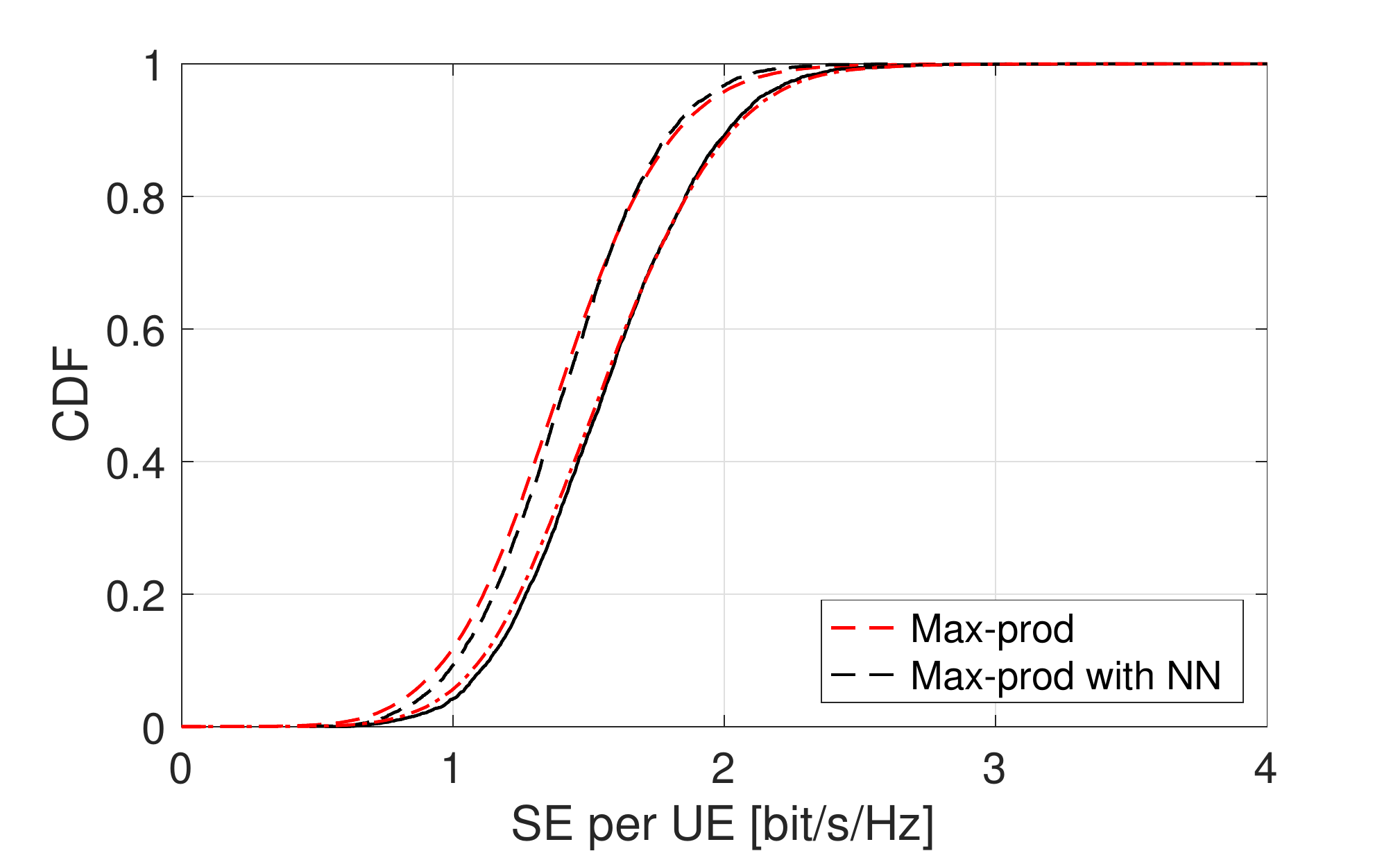}
\put(34,41){\small{MR}}
\put(46,32){\small{M-MMSE}}
\end{overpic}} 
\subfigure[CDF of the MSE]{\label{fig1b}
\begin{overpic}[unit=1mm,width=1\columnwidth]{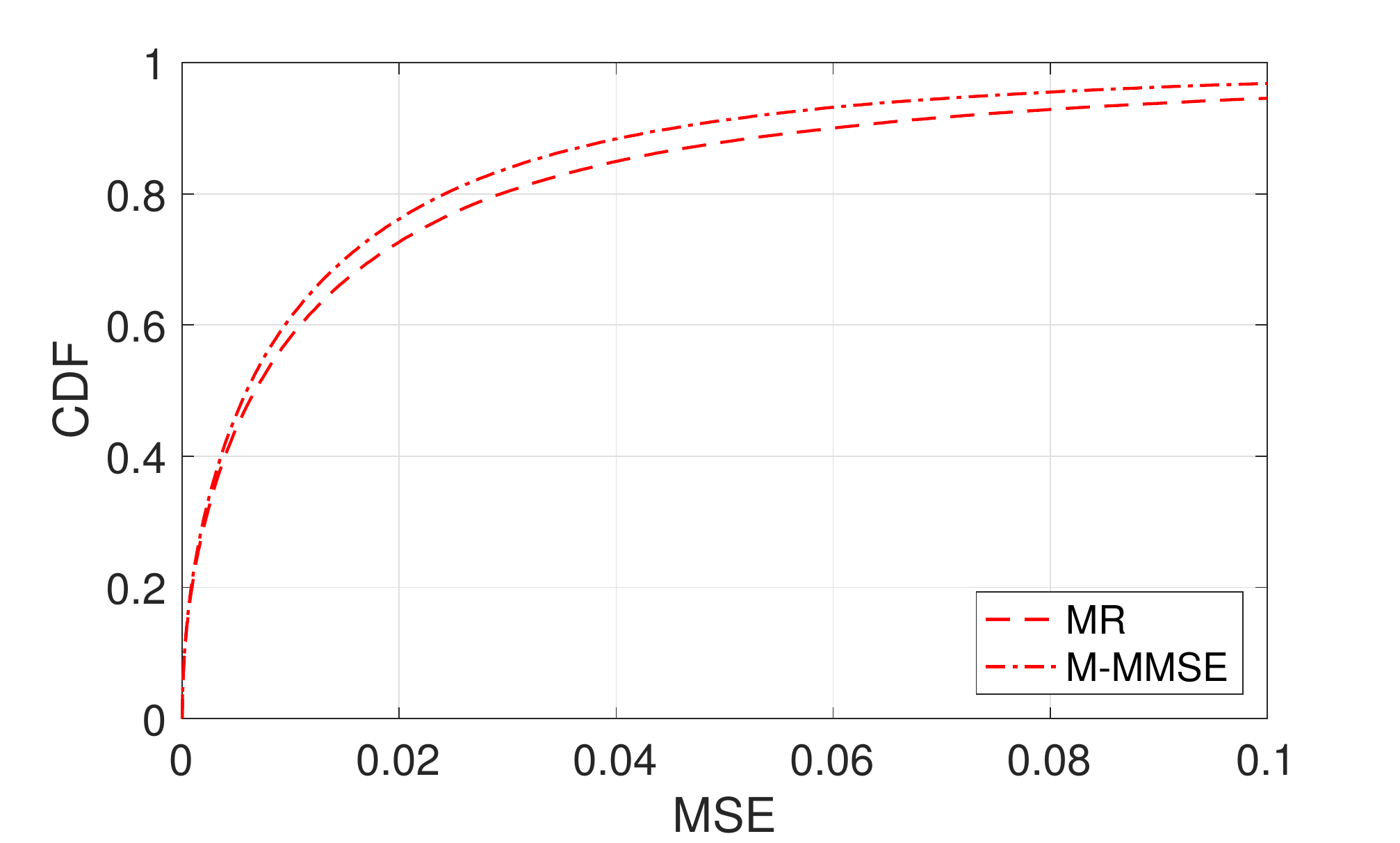}
\end{overpic}} 
\end{center} 
\caption{CDF of the DL SE with MR and M-MMSE precoding by using the neural network of Table \ref{table1}.}\label{SE_with_imperfect_statistics}
\end{figure}
\section{Conclusions}
In this work, we proposed a deep learning framework to allocate the power in the DL of a Massive MIMO network with MR and M-MMSE precoding. Two power allocation strategies were considered, namely, max-min and max-prod. We showed that with both strategies a properly trained feed-forward NN is able to learn how to allocate powers to the UEs in each cell. This is achieved by using only the knowledge of the positions of UEs in the network, thereby substantially reducing the complexity and processing time of the optimization process. Numerical results showed that the deep learning framework performs better with M-MMSE rather than with MR. This is likely due to the fact the M-MMSE allows the NN to exploit the most its available information. Moreover, the max-min policy revealed to be harder to learn. In fact, we needed to resort to recurrent neural networks with a relatively high number of trainable parameters.

The analysis was conducted for a relatively small Massive MIMO network with $L = 4$ cells and $K = 5$ UEs per cell. Further investigations are needed to understand how the developed framework performs as the size of the network increases. Moreover, in practice the number of UEs per cell varies constantly. A simple way to handle this would be to have multiple NNs per BS for all possible configurations of UEs. However, such a solution is not scalable. Besides these and many other open issues, the integration of deep learning tools for real-time power allocation in Massive MIMO seems quite promising.

\bibliographystyle{IEEEbib}
\bibliography{ref,ref_book}

\begin{thebibliography}{10}

\bibitem{Marzetta2010a}
T.~L. Marzetta,
\newblock ``Noncooperative cellular wireless with unlimited numbers of base
  station antennas,''
\newblock vol. 9, no. 11, pp. 3590--3600, 2010.

\bibitem{Larsson2014}
E.~G. Larsson, F.~Tufvesson, O.~Edfors, and T.~L. Marzetta,
\newblock ``Massive {MIMO} for next generation wireless systems,''
\newblock {\em IEEE Commun. Magazine}, vol. 52, no. 2, pp. 186--195, Feb. 2014.

\bibitem{Parkvall2017a}
S.~Parkvall, E.~Dahlman, A.~Furusk\"ar, and M.~Frenne,
\newblock ``{NR}: The new {5G} radio access technology,''
\newblock {\em IEEE Communications Standards Magazine}, vol. 1, no. 4, pp.
  24--30, Dec 2017.

\bibitem{BHS18A}
E.~Bj{\"o}rnson, J.~Hoydis, and L.~Sanguinetti,
\newblock ``Massive {MIMO} has unlimited capacity,''
\newblock {\em IEEE Transactions on Wireless Communications}, vol. 17, no. 1,
  pp. 574--590, 2018.

\bibitem{Bengio2016}
I.~Goodfellow, Y.~Bengio, and A.~Courville,
\newblock {\em Deep Learning},
\newblock MIT Press, 2016.

\bibitem{Hornik89}
K.~Hornik, M.~Stinchcombe, and H.~White,
\newblock ``Multilayer feedforward networks are universal approximators,''
\newblock {\em Neural Networks}, vol. 2, no. 5, pp. 359--366, 1989.

\bibitem{SunDL}
H.~Sun, X.~Chen, Q.~Shi, M.~Hong, X.~Fu, and N.~D. Sidiropoulos,
\newblock ``Learning to optimize: Training deep neural networks for
  interference management,''
\newblock {\em IEEE Transactions on Signal Processing}, vol. 66, no. 20, pp.
  5438--5453, 2018.

\bibitem{YuDL2018}
W.~Cui, K.~Shen, and W.~Yu,
\newblock ``Spatial deep learning for wireless scheduling,''
\newblock {\em https://arxiv.org/pdf/1808.01486.pdf}, 2018.

\bibitem{massivemimobook}
Emil Bj\"{o}rnson, Jakob Hoydis, and Luca Sanguinetti,
\newblock ``Massive {MIMO} networks: {Spectral}, energy, and hardware
  efficiency,''
\newblock {\em Foundations and Trends{\textregistered} in Signal Processing},
  vol. 11, no. 3-4, pp. 154--655, 2017.

\bibitem{Marzetta2016a}
T.~L. Marzetta, E.~G. Larsson, H.~Yang, and H.~Q. Ngo,
\newblock {\em Fundamentals of {M}assive {MIMO}},
\newblock Cambridge University Press, 2016.

\bibitem{Ngo2017a}
H.~Q. Ngo and E.~G. Larsson,
\newblock ``No downlink pilots are needed in {TDD} massive {MIMO},''
\newblock vol. 16, no. 5, pp. 2921--2935, 2017.

\bibitem{Caire17}
G.~Caire,
\newblock ``On the ergodic rate lower bounds with applications to {Massive
  MIMO},''
\newblock {\em IEEE Trans. Wireless Commun.}, vol. 17, no. 5, pp. 3258--3268,
  May 2018.

\bibitem{Li2016a}
X.~Li, E.~Bj{\"{o}}rnson, E.~G. Larsson, S.~Zhou, and J.~Wang,
\newblock ``Massive {MIMO} with multi-cell {MMSE} processing: Exploiting all
  pilots for interference suppression,''
\newblock {\em {EURASIP} J. Wirel. Commun. Netw.}, vol. 1, no. 117, 2017.

\bibitem{KingmaB14}
Diederik~P. Kingma and Jimmy Ba,
\newblock ``Adam: A method for stochastic optimization.,''
\newblock {\em CoRR}, vol. abs/1412.6980, 2014.

\end{thebibliography}

\end{document}